\documentclass[aps,twocolumn,groupedaddress,showpacs]{revtex4}
\usepackage{ifpdf}
\usepackage[dvipdfm]{graphicx}
\usepackage{epsfig,amsbsy,bm,amssymb}
\usepackage{pslatex}

\newcommand{\isum}%
{\mathop{\hbox{$\displaystyle\sum\kern-13.2pt\int\kern1.5pt$}}}
\renewcommand{\r}{{\bm r}}

\newcommand{\p}{{\bm p}}

  \newcommand{\bs}{\bigskip}
\newcommand{\bt}{\begin{tabular}}
\newcommand{\et}{\end{tabular}}

\newcommand{\eref}[1] {(\ref{#1})}
\newcommand{\Eref}[1] {Eq.~(\ref{#1})}

\newcommand{\Fref}[1] {Figure \ref{#1}}

\newcommand{\np}{\newpage}

\newcommand{\br}{\begin{eqnarray*}}
\newcommand{\er}{\end{eqnarray*}}
\newcommand{\ba}{\begin{eqnarray}}
\newcommand{\ea}{\end{eqnarray}}
\newcommand{\be}{\begin{equation}}
\newcommand{\ee}{\end{equation}}
\newcommand{\hs}{\hspace*}
\newcommand{\vs}{\vspace*}
\newcommand{\bp}{\begin{minipage}}
\newcommand{\ep}{\end{minipage}}

\begin{document}
\bibliographystyle{apsrev}

\title { Transverse electron momentum distribution in tunneling and
over the barrier ionization\\ by laser pulses with varying ellipticity}

\author{I. A. Ivanov$^{1,2}$ }
\email{Igor.Ivanov@anu.edu.au}
\author{A.S.Kheifets$^2$}

\author{J.E. Calvert$^3$}
\author{S. Goodall$^3$}
\author{X. Wang$^3$}
\author{Han Xu$^3$}
\author{A.J. Palmer$^3$}
\author{D. Kielpinski$^3$}
\author{I.V. Litvinyuk$^3$}
\author{R.T.Sang$^3$}
\email{ R.Sang@griffith.edu.au}

\affiliation{$^1$Center for Relativistic Laser Science, Institute for
Basic Science, Gwangju 500-712, Republic of Korea}

\affiliation{$^2$Research School of Physics and Engineering,
The Australian National University,
Canberra ACT 0200, Australia}

\affiliation{$^3$School of Natural Sciences and Centre for Quantum Dynamics,
Griffith University, Brisbane QLD 4111, Australia}

\date{\today}

\begin{abstract}
We study transverse electron momentum distribution (TEMD) in strong
field atomic ionization driven by laser pulses with varying
ellipticity.
We show, both experimentally and theoretically, that the TEMD in the
tunneling and over the barrier ionization regimes evolves in a
qualitatively different way when the ellipticity parameter describing
polarization state of the driving laser pulse increases.
\end{abstract}

\maketitle

A highly non-linear interaction of ultra-short light pulses with
matter enabled studying electron dynamics on the atomic time scale and
facilitated emergence of the attosecond science \cite{kri}. In
addition, strong field atomic ionization proved itself a potent tool
to interrogate atomic and molecular orbital structure via high order
harmonic radiation \cite{Shafir2009}, tunneling and diffraction
\cite{Meckel13062008} or tunneling and momentum imaging
\cite{coul7}. This utility of strong field atomic ionization is based
on the electric field of the laser pulse bending the Coulomb barrier
and letting a bound electron to tunnel out from an atom or a molecule.

Within the premise of the Keldysh theory \cite{Keldysh64}, this
tunneling regime of strong field ionization corresponds to a small
value of the adiabaticity parameter $\gamma=\omega\sqrt{2I_p}/E<1$
defined via the frequency $\omega$ and the strength $E$ of the laser
field and the ionization potential $I_p$ of the target atom (the
atomic units are used in the paper unless otherwise specified). A
finer distinction arises when one realizes that the Keldysh theory in
its original form is not applicable for very high field strengths
exceeding the over the barrier (OBI) limit.  The OBI regime was first
observed in \cite{obi1} (see also \cite{obi2} for a detailed review).
The Keldysh theory in its original form fails in the OBI regime
because there is a classical escape trajectory for an electron.  One
cannot, therefore, rely on the saddle point method that Keldysh
employed in his original work. The so-called Keldysh-Faisal-Reiss
(KFR) theory \cite{Faisal73,Reiss80} must be used instead to describe
the OBI regime.
We also note that the Keldysh theory \cite{Keldysh64} and its
subsequent developments and generalizations \cite{ppt,tunr,tunr2,adk1}
describe the quasistatic limit of small laser pulse frequencies.  In
strict terms, the Keldysh approach provides a leading-order term in
the asymptotic expansion of the ionization rate, a systematic way to
obtain higher order terms is described in \cite{tolst,tolst1}.

Despite the fact that underlying physics is very different in the two
regimes (a classically forbidden trajectory for tunneling and a
classically allowed trajectory for OBI), the energy spectra and
electron angular distributions as given by these two theories are not
dissimilar.  In this Letter, we demonstrate, that the transverse
electron momentum distribution (TEMD) is a measurable quantity that is
qualitatively different in the tunneling and the OBI regimes.  This
distribution (also known in the literature as the lateral electron
momentum distribution \cite{lat}) gives the probability to detect a
photoelectron with a given value of the momentum component $p_{\perp}$
perpendicular to the polarization plane of the laser radiation. In the
tunneling regime, TEMD exhibits a cusp-like structure due to the
Coulomb focusing effect at $p_{\perp}=0$ for linear polarization
\cite{cusp2}, and a Gaussian-like structure predicted by the Keldysh
theory for circular polarization \cite{coul7}. We studied this
transition from the cusp-like to the Gaussian structures in detail in
the tunneling regime \cite{cuspm}, and interpreted this transition as
a gradual diminishing of the role of the Coulomb effects with 
growing ellipticity of the laser pulse.  Further study of the role of
the Coulomb focusing effects was reported in \cite{cuspm1}.  We
shall see below that the situation is quite different in the OBI
regime, where the TEMD always has a cusp regardless of the value of
the ellipticity parameter. As a result of this qualitatively different
behavior of the TEMD, one can clearly distinguish the tunneling and
OBI regimes.  This is an important result since the TEMD conveys
information about the fine details of the strong field ionization
process \cite{cusp3,coul5}. One such detail is the electron velocity
distributions at the moment of time when ionization occurs, which is
often used in various models of strong field ionization. The
omnipresence of the cusp in the OBI regime also makes it unsuitable
for momentum imaging proposed in \cite{coul7}. 

As the case study, in the present work we select two markedly
different atomic species: the argon atom in the $^1S_0$ ground state
and the neon atom in the $^3P_2$ metastable state with the ionization
potentials of 15.76~eV and 5.07 eV, respectively. 
An estimate for the critical field corresponding to the onset of OBI
can be found from the equation $E_{\rm obi} \simeq I_p^2/2$ which
follows from considering the hydrogen atom placed in an external field using the parabolic
coordinates \cite{LL3}. This
rough estimate, which does not account for the above-barrier
reflection \cite{obi2} and the Stark shift, places the OBI onset of Ar
well into the $10^{14}$~W/cm$^2$ intensity range while for Ne* this
onset starts already in the $10^{12}$~W/cm$^2$ range. This comfortable
two orders of magnitude difference allows to drive these targets to
the tunneling and OBI regimes with comparable laser intensities in the
same experimental set up ( Ar@ $4.8\times 10^{14}$ W/cm$^2$ and
metastable Ne$^*$@ $2\times 10^{14}$ W/cm$^2$, both corresponding to a
similar adiabaticity parameter $\gamma\simeq0.7$ at 800~nm). To our knowledge
the only known momentum imaging experiment in similar OBI regime was
reported on Li \cite{PhysRevA.83.023413}.

A schematic representation of the experiment is shown in
\Fref{experiment}.  The ultrafast light pulses are produced by a
commercially available chirped-pulse amplification laser system
(Femtolasers, Femtopower Compact Pro CE Phase). The light pulses are
generated, stretched, amplified and then compressed in the system.
The pulse repetition rate is 1kHz with a pulse duration of ~ 6 fs,
pulse energies of approximately 450~mJ.  The pulse train is focused
down to a spot size of 7~$\mu$m radius (FWHM) at the interaction
region of the Reaction Microscope (REMI).
This is the electron detection
device, where the laser pulse ionizes atoms from a target atom beam,
then uses position dependent delay-line time of flight detectors to
determine the momentum vectors of the ionized electrons.  More
information on the experimental setup can be found in \cite{cexp1}.
The electron momentum is measured as a function of the ellipticity of
the ionizing beam, which is varied using a quarter waveplate.  The Ar
beam is provided by a cold gas jet source.  Metastable $^3P_2$ neon
atoms are produced by a gas discharge source, which uses a DC
discharge across a supersonic gas expansion region to excite
approximately 1\% of neon atoms in a gas jet into the correct state.
The flux of metastable neon atoms is improved by optical collimation
techniques that take advantage of the 640 nm closed optical transition
to the $^3D_3$ state.  Further details of this gas source can be found
in \cite{cexp2,cexp3}.

\begin{widetext}
\begin{center}
\begin{figure}[b]
\vs{-2.5cm}
\epsfxsize=11cm
\epsffile{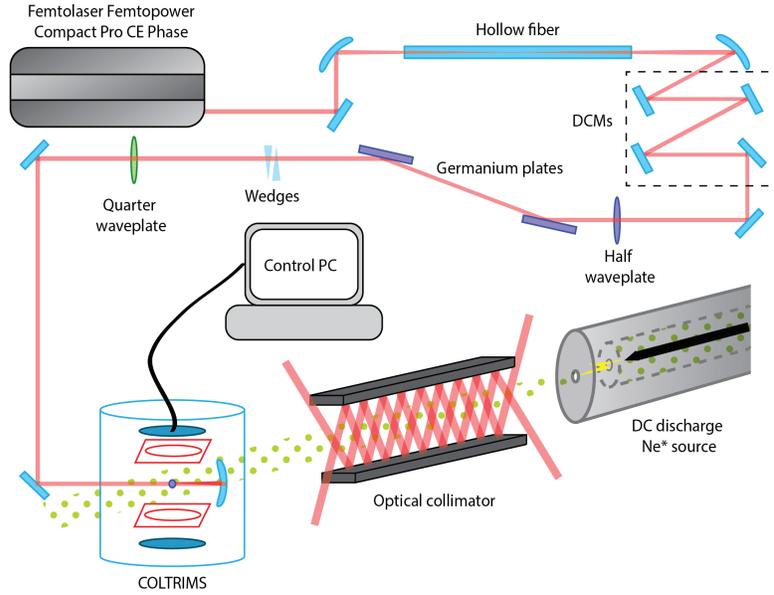} 
\vs{-.3cm}
\caption{Schematic representation of the experiment}
\label{experiment}
\end{figure}
\end{center}
\end{widetext}
\bs

Our theoretical results are obtained by solving the
time-dependent Schr\"odinger equation (TDSE):
\begin{equation}
i {\partial \Psi(\r) / \partial t}=
\left(\hat H_{\rm atom} + \hat H_{\rm int}(t)\right)
\Psi(\r) \ .
 \label{tdse}
\end{equation}
  To describe the field-free Ar and metastable Ne$^*$ atoms, we used
effective one-electron potentials \cite{oep}. The interaction of the
atom with the laser pulse is described in the velocity form of the
interaction operator:
\be
\label{gauge}
\hat H_{\rm int}(t) =
 {\bm A}(t)\cdot \hat{\bm p} \ ,
\ee
where ${\bm A}(t)$ is the vector potential of the laser pulse.
The laser pulse is elliptically polarized and propagates along the
$z$-direction which is assumed to be the quantization axis:
\be
E_x= {E\over \sqrt{1+\epsilon^2}} f(t) \cos{\omega t} \ \ , \ \
E_y= {E\epsilon\over \sqrt{1+\epsilon^2}} f(t) \sin{\omega t} \ ,
\label{ef}
\ee
where $\epsilon$ is the ellipticity parameter.  The function $f(t)=
\sin^2(\pi t/ T_1)$, with $T_1$ being the total pulse duration, is
used to represent the pulse envelope.  For the Ar atom, the field
strength was $E=0.1171$~a.u. corresponding to the experimental peak
intensity of $4.8\times 10^{14}$ W/cm$^2$.  For the metastable Ne$^*$
atom, $E=0.0756$~a.u.  with the peak intensity of $2\times 10^{14}$
W/cm$^2$.  The carrier wavelength $\lambda=800$~nm and the FWHM of
6~fs were the same for Ar and Ne$^*$.
 
To solve the TDSE we employ the strategy used in the previous works
\cite{atim,dstrong,cuspm}.
The solution of the TDSE is represented as a partial waves series:
\be
\Psi({\bm r},t)=
\sum\limits_{l=0}^{L_{\rm max}} \sum\limits_{\mu=-l}^{l}
f_{l\mu}(r,t) Y_{l\mu}(\theta,\phi).
\label{basis}
\ee
The radial part of the TDSE is discretized on the grid with the
stepsize $\delta r=0.1$~a.u. in a box of the size $R_{\rm
max}=400$~a.u.  The maximum orbital momentum in \Eref{basis} was
restricted to $L_{\rm max}=60$.  Convergence with respect to variation
of $\delta r$, $R_{\rm max}$ and $L_{\rm max}$ was carefully
monitored. The matrix iteration method \cite{velocity1} was used to
propagate TDSE in time.
Ionization amplitudes $a(\p)$ were obtained by projecting the solution
of the TDSE after the end of the pulse on the set of the ingoing
scattering states $\Psi^-_{\p}(\r)$ of the target atom.  The TEMD
$W(p_{\perp})$ was calculated as 
\be
W(p_{\perp})=\int |a(\p)|^2\ dp_x\ dp_y
 \ , \ \ \ 
p_{\perp}\equiv p_z
\label{wp}
\ee

{
\begin{figure}[t]
\vs{8cm}
\hs{-1cm}
\epsfxsize=6cm
\epsffile{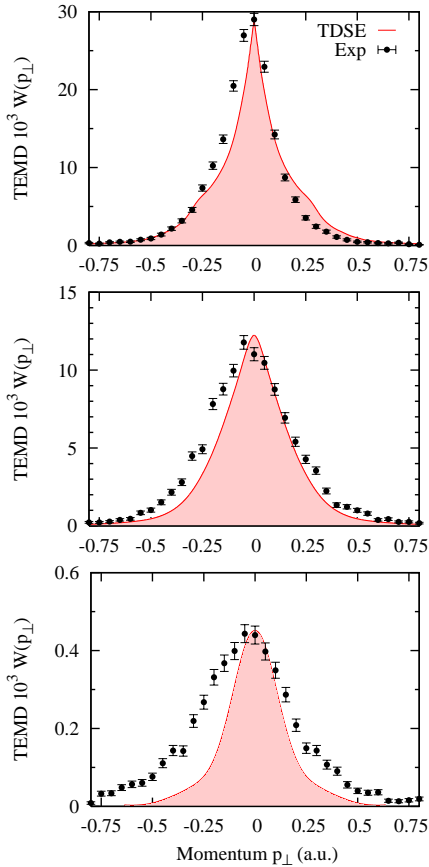} 
\caption{(Color online) TEMD of Ar (multiplied by $10^3$) for
ellipticity parameters $\epsilon=0$, 0.42, and 1 (from top to
bottom). TDSE calculation  is shown by a (red) solid curve (shaded for
a clearer  appearance), experimental data are plotted with error bars.
The peak intensity of the laser pulses is  $4.8\times 10^{14}$ W/cm$^2$. 
\label{fig1}}
\end{figure}

\begin{figure}[h]

\vs{8cm}
\hs{-1cm}
\epsfxsize=6cm
\epsffile{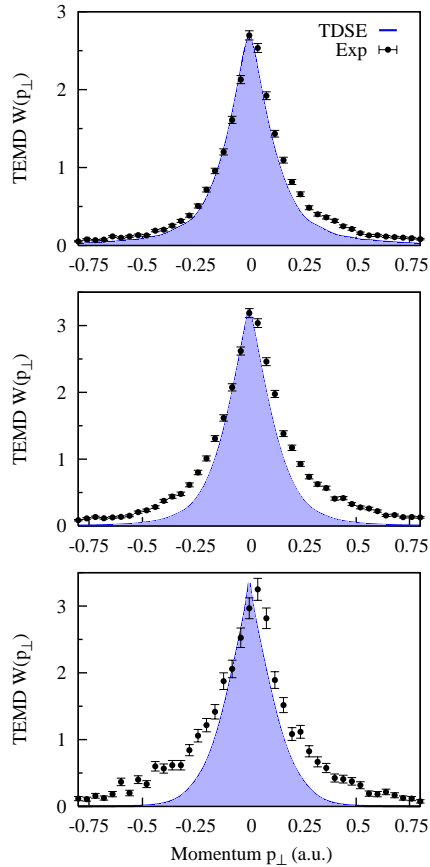} 

\caption{(Color online) TEMD of metastable Ne$^*$ for  ellipticity
parameters $\epsilon=0$, 0.42, and 1 (top to bottom). 
TDSE calculation  is shown by a (blue) solid curve (shaded for
a clearer  appearance), experimental data are plotted with error bars.
The peak intensity of the laser pulses is  $2\times 10^{14}$ W/cm$^2$. 
\label{fig2}}
\end{figure}

Experimental and theoretical TEMD results for Ar are shown in
\Fref{fig1}. A general trend of the calculated TEMD with increase of
ellipticity is very similar to that reported previously for the
simulations of the hydrogen atom \cite{cuspm}. The cusp-like structure
is present for linear polarization and it gradually evolves into a
Gaussian distribution as the ellipticity parameter
increases. Agreement between the theory and experiment is good for
linear polarization but gradually deteriorate with an increase of
ellipticity. 

\Fref{fig2} presents the theoretical and experimental TEMD results for
metastable Ne$^*$.  In this target atom, the TEMD evolution with the
ellipticity parameter is greatly reduced with the cusp clearly present
even for the circularly polarized pulse.  Similar to Ar, agreement
between the theory and experiment progressively worsens from the top
to bottom panels.

To analyze the cusp in more detail, we zoomed in on the narrow range
of momenta $|p_{\perp}|\le 0.25$~ a.u. and analyzed the function
$V(p_{\perp})=\ln{W(p_{\perp})}$ in this interval. For the TEMD
$W(p_{\perp})$ to have a cusp, $V(p_{\perp})$ should have an infinite
derivative of some order and have an expansion near $p_{\perp}=0$:
\be
V(p_{\perp})= B+A|p_{\perp}|^\alpha\ .
\label{exp}
\ee
Such expansions, in fact, reproduce $V(p_{\perp}\simeq0)$ very
well as was shown in \cite{cuspm}.

\begin{figure}[h]

\vspace*{3.5cm}
\epsfxsize=6cm
\epsffile{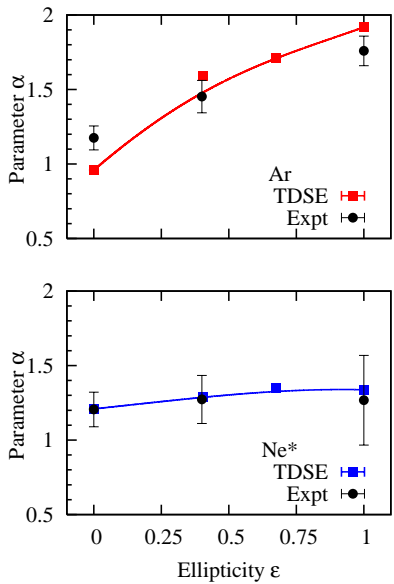} 

\caption{(Color online) The fitting parameter $\alpha$ in \Eref{exp}
as a function of the ellipticity parameter $\epsilon$ for Ar (top) and
Ne$^*$ (bottom).  TDSE results are shown with squares (a smooth solid
line is to guide the eye). The experimental data points are shown with
error bars.
\label{fig4}}
\end{figure}

The same functional form \eref{exp} was used to fit both the
theoretical and experimental data for the ground state Ar  and the
metastable Ne$^*$  in the whole range of ellipticities by
considering the coefficients $A$, $B$, $\alpha$ as fitting
parameters. The most essential $\alpha$ parameters are  shown in
\Fref{fig4} for Ar (top) and Ne$^*$ (bottom).
Both theoretical and experimental values are shown with error bars
which represent the fitting error. For the theoretical data,
this error does not exceed a fraction of a percent and is not
visible on the scale of the figure. 

The $\alpha$ parameters shown on the top and bottom panels of
\Fref{fig4} demonstrate a qualitatively different behavior as
functions of the ellipticity. For the Ar atom, the $\alpha$ parameter
grows with $\epsilon$ reaching the value close to 2 for circular
polarization. This implies that TEMD becomes close to a Gaussian
$ W(p_\perp) \propto \exp\left( -p_\perp^2/\sigma^2 \right) $
with the Gaussian width  related to the fitting parameter
$
\sigma = 1/\sqrt{-A} \ .
$
The corresponding numerical values of $0.25\pm0.002$ and $0.28\pm0.02$
for the TDSE and experiment, respectively, are close to the
experimental values reported in \cite{coul7} for comparable field
intensities.
In the meantime, the $\alpha$ parameters for the metastable Ne$^*$
atom remain essentially flat, indicating that a cusp-like behavior is
present for all $\epsilon$ in the range from linear to circular
polarization. In this case, extraction of the Gaussian width parameter
is not possible even for the circular polarization. Several TDSE
calculations performed for different field intensities did not show
any considerable variation of the cusp width. However, the
Gaussian width varies with the field strength as $\sigma \propto
E^{1/2}$. This may explain, at least partially, deviation between the
measured and calculated TEMD with circular polarization due to the
variation of the field strength across the laser-atom interaction
region while the calculation was performed at a single nominal field
intensity.  

While the Ar case shows the behavior qualitatively similar to that
found previously for hydrogen \cite{cuspm}, the metastable Ne$^*$
presents a different trend, with the cusp never disappearing
completely. In this case, a simplified description based on the
Keldysh theory is never correct even qualitatively.  This qualitative
difference can be explained by the different ionization regimes for Ar
and metastable Ne$^*$.

\begin{figure}[t]

\vspace*{4cm}
\epsfxsize=5.5cm
\epsffile{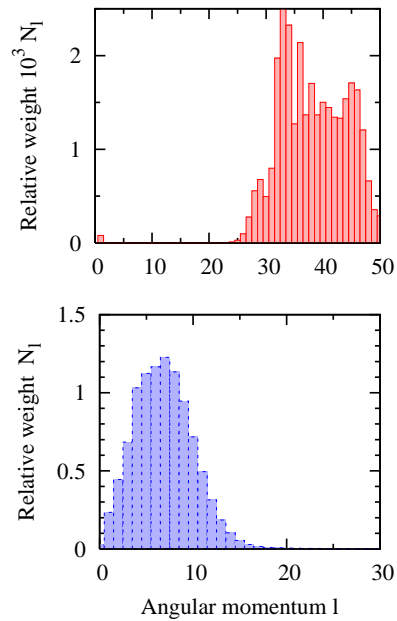} 

\caption{(Color online) Angular momentum distribution $N_l$ for
  Ar@ $4.8\times 10^{14}$ W/cm$^2$ (top) and metastable Ne$^*$@
  $2\times 10^{14}$ W/cm$^2$, bottom). Laser field is circularly
  polarized.
\label{fig5}}
\end{figure}

The TEMD cusp disappearance with increasing $\epsilon$ can be related
to a dramatic change of the angular momentum composition of the
ionized electron wave function \cite{cuspm}. This composition is
characterized by the distribution of the norm $N_l$ of the wave
function obtained if only the terms with spherical harmonics of rank
$l$ are retained in expansion \eref{basis}. For a tunneling process
this distribution is shifted towards larger $l$ with increasing pulse
ellipticity parameter.  Indeed, tunneling can be viewed as a
non-resonant absorption of a large number of photons. Absorption of a
photon from the circularly polarized wave increases the magnetic
quantum number by one unit.  This leads to a prevalence of high
angular momenta in the partial wave expansion \eref{basis}. High
angular momenta create large centrifugal barrier preventing recolliding electron 
trajectories, thereby suppressing the
Coulomb focusing effects. The cusp, therefore, vanishes for
polarization close to circular, as in the case of Ar reported here, or
hydrogen \cite{cuspm}. The situation with the metastable Ne$^*$ is
completely different. OBI dominates in this case, and since OBI is
essentially a distortion of the atomic potential to the degree, that
there is effectively a zero barrier to the continuum, the atom does
not have to absorb many photons to become ionized.  The distribution
$N_l$, therefore, is peaked at lower values of the angular momenta.
That this is indeed the case can be seen in \Fref{fig5}, where the
distributions $N_l$ are shown for Ar for the intensity of $4.8\times
10^{14}$ W/cm$^2$ (tunneling) and Ne$^*$ for the field intensity of
$2\times 10^{14}$ W/cm$^2$ (the OBI regime).  Smaller angular momenta
enhance the area near the origin where the Coulomb focusing effect is
strongest.  Larger angular momentum components are repelled from the
origin due to the centrifugal barrier.  Hence in the former case the
cusp is always present whereas in the latter case it gradually
vanishes.  This corresponds to a classical trajectory starting from
the origin whereas a tunneling trajectory starts at the point of exit
from the tunnel. In the OBI regime the electron's classical trajectory starts at the ion core
regardless of the polarization of the laser pulse, 
which may be enough for the 
efficient Coulomb focusing even if the trajectory never returns to the core.

To summarize, we described an effect of bending the Coulomb barrier of
the atom on the transverse electron momentum distribution (TEMD) in
strong field ionization in the tunneling regime. This fundamental
effect, which should be present in any atomic or molecular target, is
measured experimentally and modeled theoretically in two markedly
different species: the ground state Ar and metastable Ne*. The effect
is substantial, it has never been described or observed before and it
enables a clear distinction between the tunneling and OBI regimes in
the experiment. Also, it has to be taken into account when using TEMD
data to interrogate electronic orbitals of the target.  

Finally, we note that the cusp disappearance in the case of circular
polarization may seem to follow from a classical
consideration. Indeed, in the circularly polarized field, the two
orthogonal field components drive the photoelectron away from the
ionized core thus reducing the Coulomb focusing effect. This classical
consideration, however, fails to explain qualitatively different TEMD
behavior in the tunneling and OBI regimes observed in the present
study.

We acknowledge support of the Australian Research Council in the form
of the Discovery Project DP120101805 and DP110101894. 
Resources of the National 
Computational Infrastructure (NCI) Facility were employed.
JEC was supported by an Australian Postgraduate Award.

\np

\begin{thebibliography}{30}
\expandafter\ifx\csname natexlab\endcsname\relax\def\natexlab#1{#1}\fi
\expandafter\ifx\csname bibnamefont\endcsname\relax
  \def\bibnamefont#1{#1}\fi
\expandafter\ifx\csname bibfnamefont\endcsname\relax
  \def\bibfnamefont#1{#1}\fi
\expandafter\ifx\csname citenamefont\endcsname\relax
  \def\citenamefont#1{#1}\fi
\expandafter\ifx\csname url\endcsname\relax
  \def\url#1{\texttt{#1}}\fi
\expandafter\ifx\csname urlprefix\endcsname\relax\def\urlprefix{URL }\fi
\providecommand{\bibinfo}[2]{#2}
\providecommand{\eprint}[2][]{\url{#2}}

\bibitem[{\citenamefont{Krausz and Ivanov}(2009)}]{kri}
\bibinfo{author}{\bibfnamefont{F.}~\bibnamefont{Krausz}} \bibnamefont{and}
  \bibinfo{author}{\bibfnamefont{M.}~\bibnamefont{Ivanov}},
  \bibinfo{journal}{Rev. Mod. Phys.} \textbf{\bibinfo{volume}{81}},
  \bibinfo{pages}{163} (\bibinfo{year}{2009}).

\bibitem[{\citenamefont{Shafir et~al.}(2009)\citenamefont{Shafir, Mairesse,
  Villeneuve, Corkum, and Dudovich}}]{Shafir2009}
\bibinfo{author}{\bibfnamefont{D.}~\bibnamefont{Shafir}},
  \bibinfo{author}{\bibfnamefont{Y.}~\bibnamefont{Mairesse}},
  \bibinfo{author}{\bibfnamefont{D.~M.} \bibnamefont{Villeneuve}},
  \bibinfo{author}{\bibfnamefont{P.~B.} \bibnamefont{Corkum}},
  \bibnamefont{and} \bibinfo{author}{\bibfnamefont{N.}~\bibnamefont{Dudovich}},
  \bibinfo{journal}{Nat. Phys.} \textbf{\bibinfo{volume}{5}},
  \bibinfo{pages}{412} (\bibinfo{year}{2009}).

\bibitem[{\citenamefont{Meckel et~al.}(2008)\citenamefont{Meckel, Comtois,
  Zeidler, Staudte, Pavici\'c., Bandulet, P\'epin, Kieffer, D\"orner,
  Villeneuve et~al.}}]{Meckel13062008}
\bibinfo{author}{\bibfnamefont{M.}~\bibnamefont{Meckel}},
  \bibinfo{author}{\bibfnamefont{D.}~\bibnamefont{Comtois}},
  \bibinfo{author}{\bibfnamefont{D.}~\bibnamefont{Zeidler}},
  \bibinfo{author}{\bibfnamefont{A.}~\bibnamefont{Staudte}},
  \bibinfo{author}{\bibfnamefont{D.}~\bibnamefont{Pavici\'c.}},
  \bibinfo{author}{\bibfnamefont{H.~C.} \bibnamefont{Bandulet}},
  \bibinfo{author}{\bibfnamefont{H.}~\bibnamefont{P\'epin}},
  \bibinfo{author}{\bibfnamefont{J.~C.} \bibnamefont{Kieffer}},
  \bibinfo{author}{\bibfnamefont{R.}~\bibnamefont{D\"orner}},
  \bibinfo{author}{\bibfnamefont{D.~M.} \bibnamefont{Villeneuve}},
  \bibnamefont{et~al.}, \bibinfo{journal}{Science}
  \textbf{\bibinfo{volume}{320}}(\bibinfo{number}{5882}), \bibinfo{pages}{1478}
  (\bibinfo{year}{2008}).

\bibitem[{\citenamefont{Arissian et~al.}(2010)\citenamefont{Arissian, Smeenk,
  Turner, Trallero, Sokolov, Villeneuve, Staudte, and Corkum}}]{coul7}
\bibinfo{author}{\bibfnamefont{L.}~\bibnamefont{Arissian}},
  \bibinfo{author}{\bibfnamefont{C.}~\bibnamefont{Smeenk}},
  \bibinfo{author}{\bibfnamefont{F.}~\bibnamefont{Turner}},
  \bibinfo{author}{\bibfnamefont{C.}~\bibnamefont{Trallero}},
  \bibinfo{author}{\bibfnamefont{A.~V.} \bibnamefont{Sokolov}},
  \bibinfo{author}{\bibfnamefont{D.~M.} \bibnamefont{Villeneuve}},
  \bibinfo{author}{\bibfnamefont{A.}~\bibnamefont{Staudte}}, \bibnamefont{and}
  \bibinfo{author}{\bibfnamefont{P.~B.} \bibnamefont{Corkum}},
  \bibinfo{journal}{Phys. Rev. Lett.} \textbf{\bibinfo{volume}{105}},
  \bibinfo{pages}{133002} (\bibinfo{year}{2010}).

\bibitem[{\citenamefont{Keldysh}(1965)}]{Keldysh64}
\bibinfo{author}{\bibfnamefont{L.~V.} \bibnamefont{Keldysh}},
  \bibinfo{journal}{Sov. Phys. -JETP} \textbf{\bibinfo{volume}{20}},
  \bibinfo{pages}{1307} (\bibinfo{year}{1965}).

\bibitem[{\citenamefont{Augst et~al.}(1991)\citenamefont{Augst, Meyerhofer,
  Strickland, and Chin}}]{obi1}
\bibinfo{author}{\bibfnamefont{S.}~\bibnamefont{Augst}},
  \bibinfo{author}{\bibfnamefont{D.~D.} \bibnamefont{Meyerhofer}},
  \bibinfo{author}{\bibfnamefont{D.}~\bibnamefont{Strickland}},
  \bibnamefont{and} \bibinfo{author}{\bibfnamefont{S.~L.} \bibnamefont{Chin}},
  \bibinfo{journal}{J. Opt. Soc. Am. B} \textbf{\bibinfo{volume}{8}},
  \bibinfo{pages}{858} (\bibinfo{year}{1991}).

\bibitem[{\citenamefont{Delone and Krainov}(1998)}]{obi2}
\bibinfo{author}{\bibfnamefont{N.~B.} \bibnamefont{Delone}} \bibnamefont{and}
  \bibinfo{author}{\bibfnamefont{V.~P.} \bibnamefont{Krainov}},
  \bibinfo{journal}{Physics-Uspekhi} \textbf{\bibinfo{volume}{41}},
  \bibinfo{pages}{469} (\bibinfo{year}{1998}).

\bibitem[{\citenamefont{Faisal}(1973)}]{Faisal73}
\bibinfo{author}{\bibfnamefont{F.~H.~M.} \bibnamefont{Faisal}},
  \bibinfo{journal}{J.~Phys.~B} \textbf{\bibinfo{volume}{6}},
  \bibinfo{pages}{L89} (\bibinfo{year}{1973}).

\bibitem[{\citenamefont{Reiss}(1980)}]{Reiss80}
\bibinfo{author}{\bibfnamefont{H.~R.} \bibnamefont{Reiss}},
  \bibinfo{journal}{Phys.~Rev.~A} \textbf{\bibinfo{volume}{22}},
  \bibinfo{pages}{1786} (\bibinfo{year}{1980}).

\bibitem[{\citenamefont{Perelomov et~al.}(1966)\citenamefont{Perelomov, Popov,
  and Terentiev}}]{ppt}
\bibinfo{author}{\bibfnamefont{A.~M.} \bibnamefont{Perelomov}},
  \bibinfo{author}{\bibfnamefont{V.~S.} \bibnamefont{Popov}}, \bibnamefont{and}
  \bibinfo{author}{\bibfnamefont{M.~V.} \bibnamefont{Terentiev}},
  \bibinfo{journal}{Sov. Phys. -JETP} p. \bibinfo{pages}{924}
  (\bibinfo{year}{1966}).

\bibitem[{\citenamefont{Popov}(2004)}]{tunr}
\bibinfo{author}{\bibfnamefont{V.~S.} \bibnamefont{Popov}},
  \bibinfo{journal}{Physics-Uspekhi} \textbf{\bibinfo{volume}{47}},
  \bibinfo{pages}{855} (\bibinfo{year}{2004}).

\bibitem[{\citenamefont{Popruzhenko}(2014)}]{tunr2}
\bibinfo{author}{\bibfnamefont{S.~V.} \bibnamefont{Popruzhenko}},
  \bibinfo{journal}{J.~Phys.~B}
  \textbf{\bibinfo{volume}{47}}(\bibinfo{number}{20}), \bibinfo{pages}{204001}
  (\bibinfo{year}{2014}).

\bibitem[{\citenamefont{Ammosov et~al.}(1986)\citenamefont{Ammosov, Delone, and
  Krainov}}]{adk1}
\bibinfo{author}{\bibfnamefont{M.~V.} \bibnamefont{Ammosov}},
  \bibinfo{author}{\bibfnamefont{N.~B.} \bibnamefont{Delone}},
  \bibnamefont{and} \bibinfo{author}{\bibfnamefont{V.~P.}
  \bibnamefont{Krainov}}, \bibinfo{journal}{Sov. Phys. -JETP} p.
  \bibinfo{pages}{1191} (\bibinfo{year}{1986}).

\bibitem[{\citenamefont{Tolstikhin and Morishita}(2012)}]{tolst}
\bibinfo{author}{\bibfnamefont{O.~I.} \bibnamefont{Tolstikhin}}
  \bibnamefont{and}
  \bibinfo{author}{\bibfnamefont{T.}~\bibnamefont{Morishita}},
  \bibinfo{journal}{Phys. Rev. A} \textbf{\bibinfo{volume}{86}},
  \bibinfo{pages}{043417} (\bibinfo{year}{2012}).

\bibitem[{\citenamefont{Trinh et~al.}(2015)\citenamefont{Trinh, Tolstikhin, and
  Morishita}}]{tolst1}
\bibinfo{author}{\bibfnamefont{V.~H.} \bibnamefont{Trinh}},
  \bibinfo{author}{\bibfnamefont{O.~I.} \bibnamefont{Tolstikhin}},
  \bibnamefont{and}
  \bibinfo{author}{\bibfnamefont{T.}~\bibnamefont{Morishita}},
  \bibinfo{journal}{J.~Phys.~B} \textbf{\bibinfo{volume}{48}},
  \bibinfo{pages}{061003} (\bibinfo{year}{2015}).

\bibitem[{\citenamefont{I.Petersen et~al.}(2015)\citenamefont{I.Petersen,
  J.Henkel, and M.Lein}}]{lat}
\bibinfo{author}{\bibnamefont{I.Petersen}},
  \bibinfo{author}{\bibnamefont{J.Henkel}}, \bibnamefont{and}
  \bibinfo{author}{\bibnamefont{M.Lein}}, \bibinfo{journal}{Phys.~Rev.~Lett.}
  \textbf{\bibinfo{volume}{114}}, \bibinfo{pages}{103004}
  (\bibinfo{year}{2015}).

\bibitem[{\citenamefont{Rudenko et~al.}(2005)\citenamefont{Rudenko, Zrost,
  Ergler, Voitkiv, Najjari, de~Jesus, Feuerstein, Schr\"oter, Moshammer, and
  Ullrich}}]{cusp2}
\bibinfo{author}{\bibfnamefont{A.}~\bibnamefont{Rudenko}},
  \bibinfo{author}{\bibfnamefont{K.}~\bibnamefont{Zrost}},
  \bibinfo{author}{\bibfnamefont{T.}~\bibnamefont{Ergler}},
  \bibinfo{author}{\bibfnamefont{A.~B.} \bibnamefont{Voitkiv}},
  \bibinfo{author}{\bibfnamefont{B.}~\bibnamefont{Najjari}},
  \bibinfo{author}{\bibfnamefont{V.~L.~B.} \bibnamefont{de~Jesus}},
  \bibinfo{author}{\bibfnamefont{B.}~\bibnamefont{Feuerstein}},
  \bibinfo{author}{\bibfnamefont{C.~D.} \bibnamefont{Schr\"oter}},
  \bibinfo{author}{\bibfnamefont{R.}~\bibnamefont{Moshammer}},
  \bibnamefont{and} \bibinfo{author}{\bibfnamefont{J.}~\bibnamefont{Ullrich}},
  \bibinfo{journal}{J.~Phys.~B}
  \textbf{\bibinfo{volume}{38}}(\bibinfo{number}{11}), \bibinfo{pages}{L191}
  (\bibinfo{year}{2005}).

\bibitem[{\citenamefont{Ivanov}(2014)}]{cuspm}
\bibinfo{author}{\bibfnamefont{I.~A.} \bibnamefont{Ivanov}},
  \bibinfo{journal}{Phys. Rev. A} \textbf{\bibinfo{volume}{90}},
  \bibinfo{pages}{013418} (\bibinfo{year}{2014}).

\bibitem[{\citenamefont{Kheifets and Ivanov}(2014)}]{cuspm1}
\bibinfo{author}{\bibfnamefont{A.~S.} \bibnamefont{Kheifets}} \bibnamefont{and}
  \bibinfo{author}{\bibfnamefont{I.~A.} \bibnamefont{Ivanov}},
  \bibinfo{journal}{Phys. Rev. A} \textbf{\bibinfo{volume}{90}},
  \bibinfo{pages}{033404} (\bibinfo{year}{2014}).

\bibitem[{\citenamefont{Pfeiffer et~al.}(2012)\citenamefont{Pfeiffer, Cirelli,
  Landsman, Smolarski, Dimitrovski, Madsen, and Keller}}]{cusp3}
\bibinfo{author}{\bibfnamefont{A.~N.} \bibnamefont{Pfeiffer}},
  \bibinfo{author}{\bibfnamefont{C.}~\bibnamefont{Cirelli}},
  \bibinfo{author}{\bibfnamefont{A.~S.} \bibnamefont{Landsman}},
  \bibinfo{author}{\bibfnamefont{M.}~\bibnamefont{Smolarski}},
  \bibinfo{author}{\bibfnamefont{D.}~\bibnamefont{Dimitrovski}},
  \bibinfo{author}{\bibfnamefont{L.~B.} \bibnamefont{Madsen}},
  \bibnamefont{and} \bibinfo{author}{\bibfnamefont{U.}~\bibnamefont{Keller}},
  \bibinfo{journal}{Phys. Rev. Lett.} \textbf{\bibinfo{volume}{109}},
  \bibinfo{pages}{083002} (\bibinfo{year}{2012}).

\bibitem[{\citenamefont{Dreissigacker and Lein}(2013)}]{coul5}
\bibinfo{author}{\bibfnamefont{I.}~\bibnamefont{Dreissigacker}}
  \bibnamefont{and} \bibinfo{author}{\bibfnamefont{M.}~\bibnamefont{Lein}},
  \bibinfo{journal}{Chemical Physics} \textbf{\bibinfo{volume}{414}},
  \bibinfo{pages}{69 } (\bibinfo{year}{2013}).

\bibitem[{\citenamefont{Landau and Lifshitz}(1965)}]{LL3}
\bibinfo{author}{\bibfnamefont{L.~D.} \bibnamefont{Landau}} \bibnamefont{and}
  \bibinfo{author}{\bibfnamefont{E.~M.} \bibnamefont{Lifshitz}},
  \emph{\bibinfo{title}{Quantum Mechanics}} (\bibinfo{publisher}{Pergamon
  Press}, \bibinfo{year}{1965}).

\bibitem[{\citenamefont{Schuricke et~al.}(2011)\citenamefont{Schuricke, Zhu,
  Steinmann, Simeonidis, Ivanov, Kheifets, Grum-Grzhimailo, Bartschat, Dorn,
  and Ullrich}}]{PhysRevA.83.023413}
\bibinfo{author}{\bibfnamefont{M.}~\bibnamefont{Schuricke}},
  \bibinfo{author}{\bibfnamefont{G.}~\bibnamefont{Zhu}},
  \bibinfo{author}{\bibfnamefont{J.}~\bibnamefont{Steinmann}},
  \bibinfo{author}{\bibfnamefont{K.}~\bibnamefont{Simeonidis}},
  \bibinfo{author}{\bibfnamefont{I.}~\bibnamefont{Ivanov}},
  \bibinfo{author}{\bibfnamefont{A.}~\bibnamefont{Kheifets}},
  \bibinfo{author}{\bibfnamefont{A.~N.} \bibnamefont{Grum-Grzhimailo}},
  \bibinfo{author}{\bibfnamefont{K.}~\bibnamefont{Bartschat}},
  \bibinfo{author}{\bibfnamefont{A.}~\bibnamefont{Dorn}}, \bibnamefont{and}
  \bibinfo{author}{\bibfnamefont{J.}~\bibnamefont{Ullrich}},
  \bibinfo{journal}{Phys. Rev. A}
  \textbf{\bibinfo{volume}{83}}(\bibinfo{number}{2}), \bibinfo{pages}{023413}
  (\bibinfo{year}{2011}).

\bibitem[{\citenamefont{Xu et~al.}(2013)\citenamefont{Xu, Maclean, Laban,
  Wallace, Kielpinski, Sang, , and Litvinyuk}}]{cexp1}
\bibinfo{author}{\bibfnamefont{H.}~\bibnamefont{Xu}},
  \bibinfo{author}{\bibfnamefont{J.-P.} \bibnamefont{Maclean}},
  \bibinfo{author}{\bibfnamefont{D.}~\bibnamefont{Laban}},
  \bibinfo{author}{\bibfnamefont{W.}~\bibnamefont{Wallace}},
  \bibinfo{author}{\bibfnamefont{D.}~\bibnamefont{Kielpinski}},
  \bibinfo{author}{\bibfnamefont{R.}~\bibnamefont{Sang}}, , \bibnamefont{and}
  \bibinfo{author}{\bibfnamefont{I.}~\bibnamefont{Litvinyuk}},
  \bibinfo{journal}{New J. Phys.} \textbf{\bibinfo{volume}{15}},
  \bibinfo{pages}{023034} (\bibinfo{year}{2013}).

\bibitem[{\citenamefont{Baker et~al.}(2003)\citenamefont{Baker, Palmer, and
  Sang}}]{cexp2}
\bibinfo{author}{\bibfnamefont{M.}~\bibnamefont{Baker}},
  \bibinfo{author}{\bibfnamefont{A.}~\bibnamefont{Palmer}}, \bibnamefont{and}
  \bibinfo{author}{\bibfnamefont{R.}~\bibnamefont{Sang}},
  \bibinfo{journal}{Meas. Sci. Technol.} \textbf{\bibinfo{volume}{14}},
  \bibinfo{pages}{N5} (\bibinfo{year}{2003}).

\bibitem[{\citenamefont{Palmer et~al.}(2004)\citenamefont{Palmer, Baker, and
  Sang}}]{cexp3}
\bibinfo{author}{\bibfnamefont{A.}~\bibnamefont{Palmer}},
  \bibinfo{author}{\bibfnamefont{M.}~\bibnamefont{Baker}}, \bibnamefont{and}
  \bibinfo{author}{\bibfnamefont{R.}~\bibnamefont{Sang}},
  \bibinfo{journal}{Rev. Sci Instrum.} \textbf{\bibinfo{volume}{75}},
  \bibinfo{pages}{5056} (\bibinfo{year}{2004}).

\bibitem[{\citenamefont{Sarsa et~al.}(2004)\citenamefont{Sarsa, G\'{a}lvez, and
  Buendia}}]{oep}
\bibinfo{author}{\bibfnamefont{A.}~\bibnamefont{Sarsa}},
  \bibinfo{author}{\bibfnamefont{F.~J.} \bibnamefont{G\'{a}lvez}},
  \bibnamefont{and} \bibinfo{author}{\bibfnamefont{E.}~\bibnamefont{Buendia}},
  \bibinfo{journal}{At. Data Nucl. Data Tables}
  \textbf{\bibinfo{volume}{88}}(\bibinfo{number}{1}), \bibinfo{pages}{163}
  (\bibinfo{year}{2004}).

\bibitem[{\citenamefont{Ivanov}(2010)}]{atim}
\bibinfo{author}{\bibfnamefont{I.~A.} \bibnamefont{Ivanov}},
  \bibinfo{journal}{Phys. Rev. A}
  \textbf{\bibinfo{volume}{82}}(\bibinfo{number}{3}), \bibinfo{pages}{033404}
  (\bibinfo{year}{2010}).

\bibitem[{\citenamefont{Ivanov}(2011)}]{dstrong}
\bibinfo{author}{\bibfnamefont{I.~A.} \bibnamefont{Ivanov}},
  \bibinfo{journal}{Phys. Rev. A}
  \textbf{\bibinfo{volume}{83}}(\bibinfo{number}{2}), \bibinfo{pages}{023421}
  (\bibinfo{year}{2011}).

\bibitem[{\citenamefont{Nurhuda and Faisal}(1999)}]{velocity1}
\bibinfo{author}{\bibfnamefont{M.}~\bibnamefont{Nurhuda}} \bibnamefont{and}
  \bibinfo{author}{\bibfnamefont{F.~H.~M.} \bibnamefont{Faisal}},
  \bibinfo{journal}{Phys. Rev. A}
  \textbf{\bibinfo{volume}{60}}(\bibinfo{number}{4}), \bibinfo{pages}{3125}
  (\bibinfo{year}{1999}).

\end{thebibliography}

\end{document}